\documentclass[a4paper,10pt,twocolumn]{article}
\usepackage{amsmath}
\usepackage{amsfonts}
\usepackage{graphicx}
\usepackage{siunitx}
\usepackage{pdfpages}

\begin{document}
%
\title{Makespan minimization of Time-Triggered traffic on a TTEthernet network}


\author{Jan Dvo\v{r}\'{a}k\\
Department of Control Engineering, FEE\\
Czech Technical University in Prague\\
Prague, Czech Republic\\
Email: dvoraj57@fel.cvut.cz
\and
Martin Heller\\
Department of Control Engineering, FEE\\
Czech Technical University\\
Prague, Czech Republic\\
Email: hellemar@alumni.fel.cvut.cz\\
\and
Zden\v{e}k Hanz\'{a}lek\\
DCE FEE and CIIRC\\
Czech Technical University in Prague\\
Prague, Czech Republic\\
Email: hanzalek@fel.cvut.cz}

\null
\includepdf[pages=1,fitpaper,noautoscale]{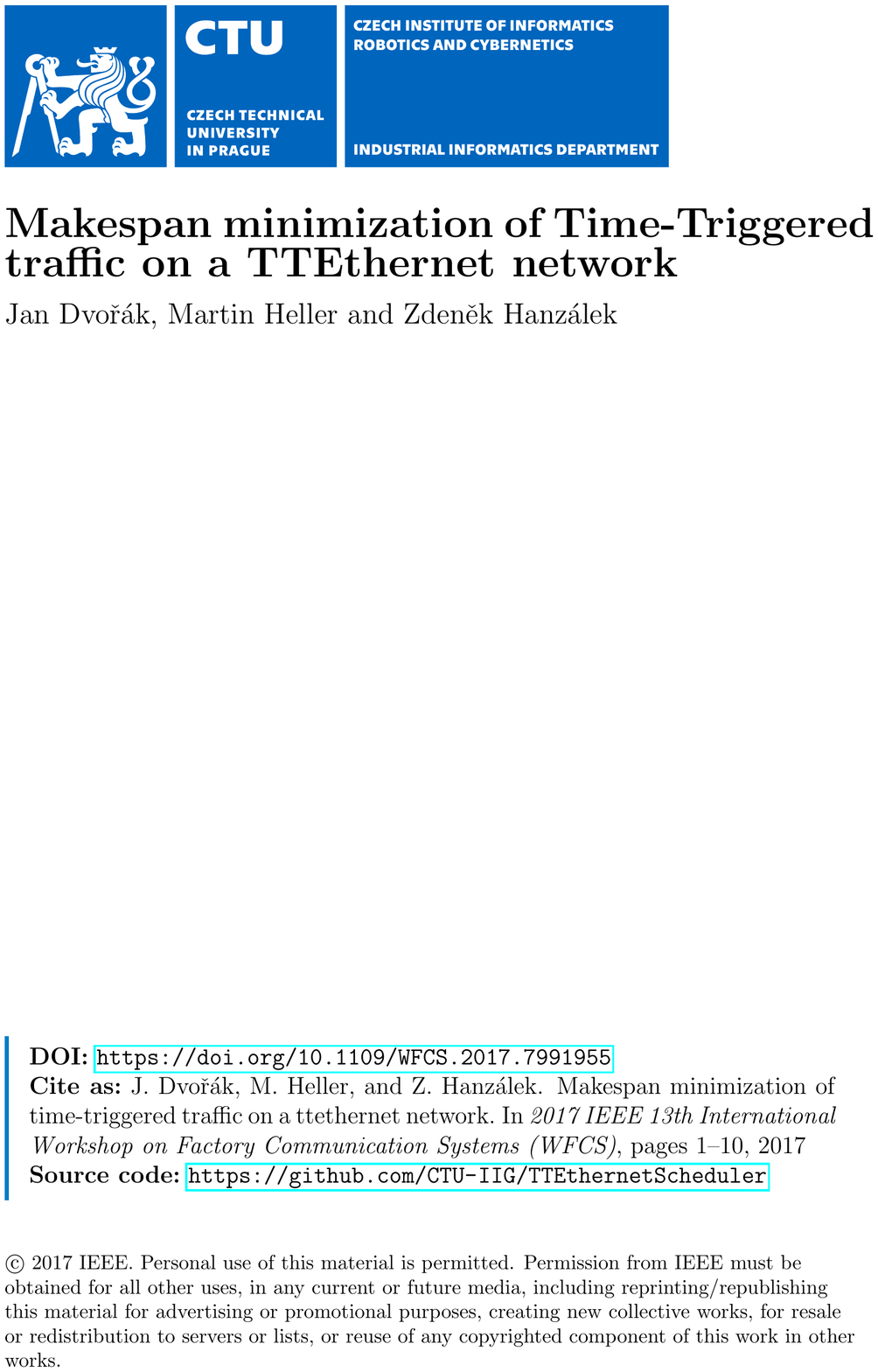}
\date{}
\maketitle

\begin{abstract}
The reliability of the increasing number of modern applications and systems strongly depends on interconnecting technology.
Complex systems which usually need to exchange, among other things, multimedia data together with safety-related information, as in the automotive or avionic industry, for example, make demands on both the high bandwidth and the deterministic behavior of the communication.
TTEthernet is a protocol that has been developed to face these requirements while providing the generous bandwidth of Ethernet up to 1\,Gbit/s and enhancing its determinism by the Time-Triggered message transmission which follows the predetermined schedule. 
Therefore, synthesizing a good schedule which meets all the real-time requirements is essential for the performance of the whole system.

In this paper, we study the concept of creating the communication schedules for the Time-Triggered traffic while minimizing its makespan.
The aim is to maximize the uninterrupted gap for remaining traffic classes in each integration cycle.
The provided scheduling algorithm, based on the Resource-Constrained Project Scheduling Problem formulation and the load balancing heuristic, obtains near-optimal (within 15\% of non-tight lower bound) solutions in 5 minutes even for industrial sized instances.
The universality of the provided method allows easily modify or extend the problem statement according to particular industrial demands.
Finally, the studied concept of makespan minimization is justified through the concept of scheduling with porosity according to the worst-case delay analysis of Event-Triggered traffic.
\end{abstract}

\section{Introduction}
This paper focuses on the problem of creating schedules for Time-Triggered traffic on the TTEthernet network. 
The objective is to minimize the makespan of the Time-Triggered traffic and, thus, maximize the continuous gap for the Event-Triggered traffic.
The aim of the paper is to develop the scheduling algorithm and, consequently, to compare the concept of the makespan minimization with the concept of the porosity optimization proposed by Steiner in~\cite{Steiner2011}.
The TTEthernet network is a network protocol which offers Time-Triggered communication and preserves the backward compatibility with Ethernet.

Ethernet (IEEE 802.3) has been the prevalent technology for home and office networks over the last few decades.
Thanks to its widespread adoption, it has developed into a mature technology offering high bandwidth with cheap and readily available hardware.

\begin{figure}[t]
\resizebox{\columnwidth}{!}
{
\includegraphics{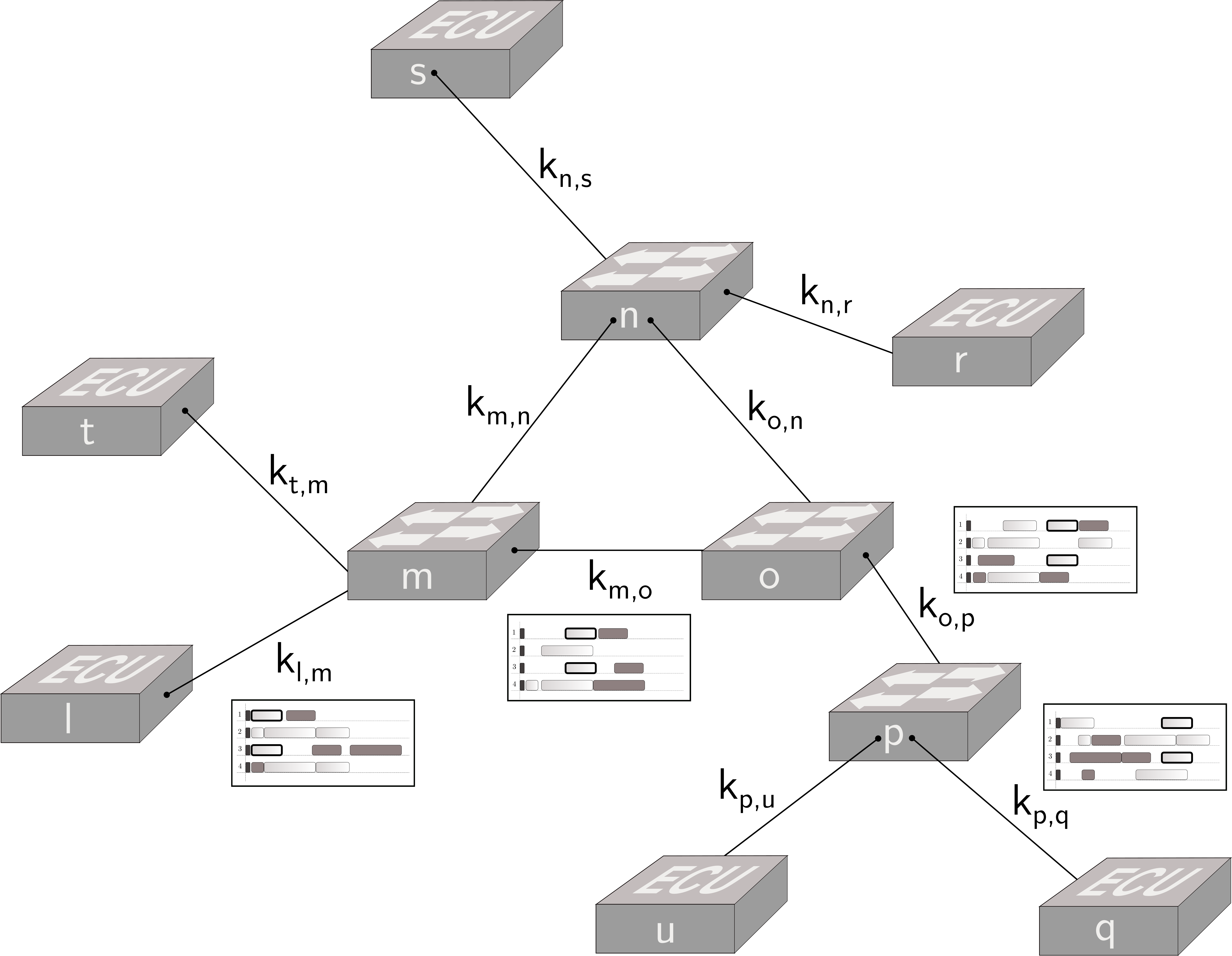}
}
\caption{Example of the TTEthernet network infrastructure with routing and scheduling of message $m_1$ from node l to node q}
\label{Fig:Infrastructure}
\end{figure}

Conventional computer networks are mainly used for on-demand data transfer without an immediate physical impact on the real world. 
In case of failure, the transmission can usually be repeated without causing major difficulties. 
Therefore, the focus is on the bandwidth and efficiency with only moderate demands on reliability.

In contrast, in industrial applications various control loops of physical devices are realized by the network, e.g. data from sensors are transferred to a processing unit which then sends commands to actuators. 
Any disturbance can, thus, have immediate effects on the real world with possible severe consequences.
As jitter (variance in transmission times) is detrimental to the function of the control loops, determinism is often required. 
In addition, individual devices in the network are often limited in hardware and need to operate in demanding environments.

Due to these differences, different technologies and protocols were traditionally used for conventional computer networks and industrial networks.
Ethernet has been used for home and office networks while various Fieldbus networks (such as CAN~\cite{Davis2013} or Profinet~\cite{Hanzalek2010}) have been used for industrial applications.
However, with the increasing integration of industrial systems, increasing demands on the volume of data transferred and also the maturing and development of the Ethernet, there has been a trend of using Ethernet-based networks for industrial applications as well.
This trend has become even more pronounced with the ever increasing amounts of data transfers necessary to facilitate features like real-time image processing and recognition or communication among individual units in a smart system.
Therefore, extensions of Ethernet are being developed to meet the demands of industrial applications.
An overview of the development of industrial networks is given by \cite{Galloway:Intro}.

TTEthernet is a promising extension of Ethernet, which provides determinism and fault-tolerance while being compatible with standard Ethernet.
Besides the TTEthernet standard, there has been ongoing effort on the standardization of extensions to Ethernet for scheduled traffic by the IEEE 802.1 Time-Sensitive Networking Task Group.

Determinism with the strictest guarantees is achieved through a fixed schedule for the traffic.
Therefore, synthesizing a good (exactly what this means will be discussed later) schedule which meets all the requirements and deadlines is essential for the performance of the network.
Because TTEthernet allows for complex topologies, the scheduling involves additional complexity compared to bus or passive star topologies of networks like FlexRay~\cite{ISOFlexRay} or CAN.


\subsection{TTEthernet Overview}
TTEthernet (TT stands for Time-Triggered) is an extension of Ethernet for
deterministic communication developed as joint project among the Vienna University of Technology~\cite{kopetz2005}, TTTech and Honeywell, and
standardized as SAE~AS~6802~\cite{AS6802} in 2011.

TTEthernet operates at Level 2 of the ISO/OSI model, above the physical layer of Ethernet.
It requires a switched network with fully duplex physical links, such as Fast Ethernet physical link 100BASE-TX or Automotive Ethernet standard 100BASE-T1, so that unpredictable conflicts, while accessing a shared medium, are avoided.
An example of the TTEthernet infrastructure is depicted in Fig.~\ref{Fig:Infrastructure}.

TTEthernet specifies a protocol for clock synchronization and the rules for managing the traffic on the network.
After an initial startup phase when the clocks of the devices in the network are synchronized for the first time, the operation of TTEthernet is periodic. 
The clocks are being periodically synchronized to counter any possible clock drift when in steady operation.
This period is called the \emph{integration cycle}.
The messages which follow a deterministic schedule are also periodic.
The least common multiple of their period is called the \emph{cluster cycle}.

TTEthernet integrates traffic of different time-criticality levels into one physical network. 
There are three traffic classes in TTEthernet.
These classes, ordered by decreasing priority, are Time-Triggered (TT), Rate-Constrained (RC) and Best-Effort (BE) traffic.

\begin{figure}[t]
\resizebox{\columnwidth}{!}
{
\includegraphics{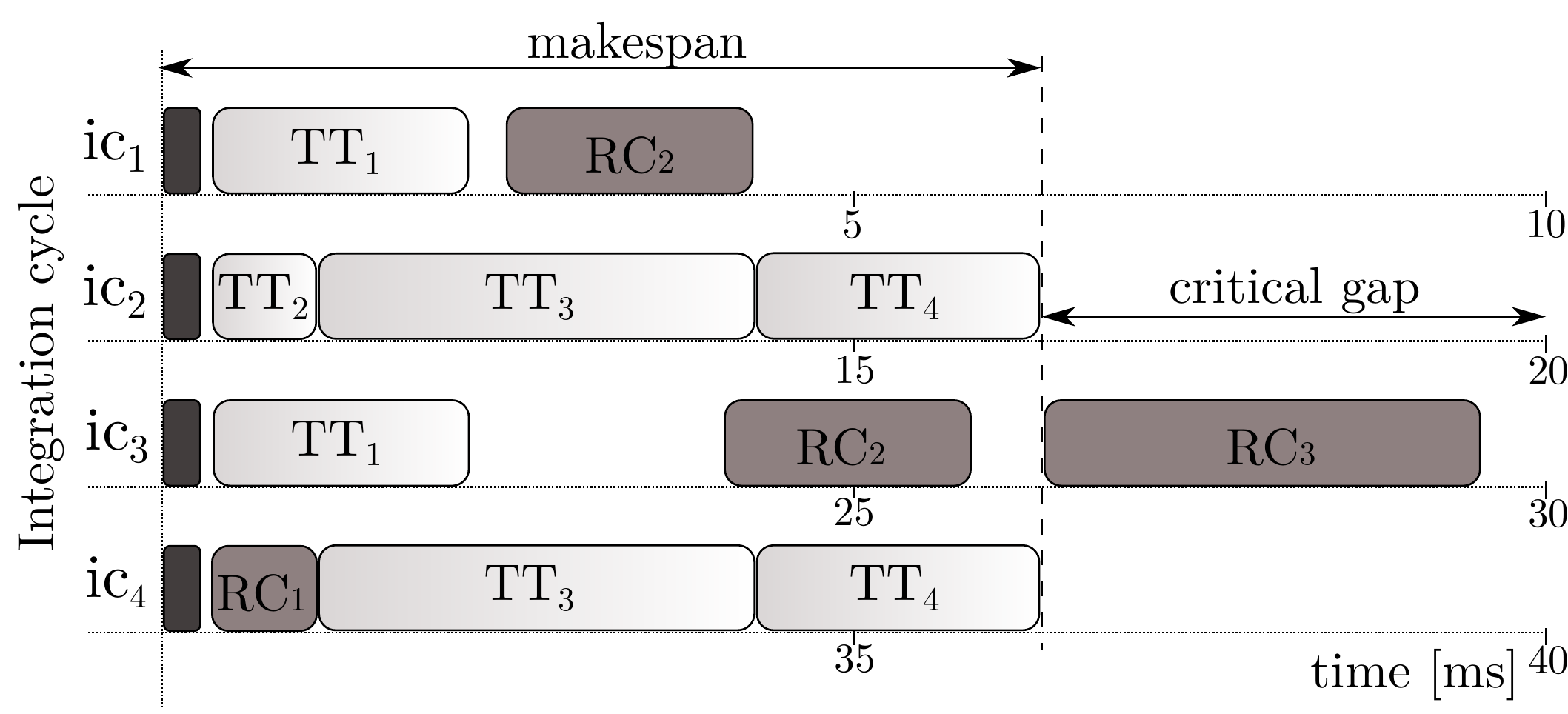}
}
\caption{Example of communication on a link in one cluster cycle}
\label{Fig:Link_schedules}
\end{figure}

The TT traffic class has the highest priority, and sub-$\mu$s jitter can be achieved (depending on the network devices).
The TT messages are periodic. 
We assume that they are strictly periodic (i.e. no jitter is allowed) in agreement with~\cite{TS2015}. 
Their schedule is calculated offline and then loaded into the individual devices.
The schedule is repetitive with a hyperperiod of the cluster cycle.

A simple example of the TT traffic together with the RC traffic on one direction of a physical link is presented in Fig.~\ref{Fig:Link_schedules}. 
In the figure, the integration cycles are situated in rows and x-axis represents the time instants in the particular integration cycle. 
The figure shows that messages $TT_1$, $TT_3$ and $TT_4$ have the same period (twice the integration cycle) and $TT_2$ has another period (four times the integration cycle). 
The length of the cluster cycle is equal to four times the length of the integration cycle.
The dark message at the beginning of each integration cycle is the synchronization message.
The synchronization message is used for periodical clock synchronization among  nodes.
This message represents the time allocated for the synchronization mechanism of the TTEthernet and it does not participate in the optimization process.

The schedule also provides temporal isolation and enables fault tolerance.
This is due to the fact that not only the sending of a frame is scheduled, but its reception is scheduled as well. 
If a TT frame arrives outside the \emph{acceptance window} (i.e. the time the frame is supposed to arrive considering the synchronization inaccuracy), it is discarded by the receiver.
This mechanism is called the \emph{temporal firewall}.

For traffic with less strict precision requirements, the RC traffic class can be used.
This traffic class conforms to the ARINC~664p7 specification \cite{ARINC664p7} (also called AFDX).
It offers greater flexibility because only the frame routing needs to be determined offline.
The messages themselves are event-driven within some limitations.

The RC traffic represents the event-triggered communication.
Thus, it does not follow any schedule known in advance.
It is organized in so-called \emph{virtual links}.
As stated in \cite{AFDXdetailed}, a virtual link is an analogy to the ARINC~429~\cite{ARINC429} single-source multi-drop bus.
The virtual link determines the routing of the messages associated with it.
Furthermore, there are two parameters: the maximum allowed frame size and the bandwidth allocation gap, associated with each virtual link.
The bandwidth allocation gap represents the minimum allowed length of an interval between consecutive frames on the virtual link.
This effectively limits the bandwidth of the virtual link.
In return for this limitation, the maximum possible delay of any RC message can be calculated offline.

Standard Ethernet traffic can also be transmitted through the network. 
Even standard Ethernet devices, unaware of TTEthernet, can communicate through the network.
Such traffic is called the Best-Effort (BE) traffic and has the lowest priority.

When the TT traffic is used together with other traffic classes, a TT frame could be delayed by another RC or BE frame.
This happens when a TT frame arrives while an RC or BE frame is in transmission.
Then, unless the transmission of the RC or BE frame can be interrupted, the TT frame has to wait until the transmission is finished.
Therefore, a method of handling such situations, called the \emph{traffic integration policy}, is needed.
There are three policies for the integration of different traffic classes as mentioned by \cite{TS2012}: Pre-emption, Shuffling and Timely block.
The Timely block integration policy, which causes no extra delay of the TT traffic, is used in this paper.
In this case, an RC frame can only be transmitted if there is enough time for the transmission of the entire frame before the next TT frame is scheduled.
If there is insufficient time, the transmission of the RC frame is postponed until after the TT frame is transmitted.
It additionally means, that the TT traffic follows the schedule without any delays.

\subsection{Related works}
Many recent publications focus on the scheduling of Time-Triggered communication on the Ethernet network.
The most significant research in this area was done by Steiner~et~al.
Their first paper~\cite{Steiner2010} described the basic constraints for scheduling the TT traffic.
The scheduling problem was solved by the reformulation of the constraints to the SMT (Satisfiability modulo theories) model.
The native SMT formulation was used just for small cases with up to 100 message instances.
The second approach presented allows one to find a schedule for bigger instances.
It uses the SMT solver as a backend for scheduling smaller groups of frames, into which the frames are divided.
After a group of frames is scheduled, the positions of the corresponding frame instances are fixed, and the next group is scheduled iteratively.
The paper does not consider other traffic classes, and the only criterion is to find any feasible schedule.
The SMT formulation of the Time-Trigged message scheduling problem has been adapted to 802.1Qbv~\cite{Craciunas2016} in 2016.
In~\cite{Steiner2011}, which builds on and extends~\cite{Steiner2010}, traffic from other classes than just the TT class is considered. 
To reserve capacity for rate-constrained traffic, the concept of schedule porosity, i.e. inserting blank slots reserved for the RC traffic into the schedule, is introduced.
The pessimistic time analysis for the RC traffic is proposed to evaluate the concept.
Tamas-Selicean~et~al. published a new method for calculating the worst-case delay~\cite{TS2015} which promises much tighter estimates compared to the previous work.
The improved precision comes, however, at the cost of much more expensive computation.
As noted by~\cite{Steiner:NewHorizons}, porosity scheduling has the disadvantage that gaps introduced at the beginning of the scheduling process do not consider the profile of the RC traffic.

The idea of using the metaheuristic approach for scheduling the TTEthernet communication by TabuSearch was proposed in~\cite{TS2012} and further extended and more intensely evaluated in~\cite{TS2015designopt}.
The operators of the TabuSearch~\cite{Baghel2012} algorithm take the RC worst-case delay calculation into account and try to change the current schedule in such a way that all the real-time constraints imposed on the RC traffic are satisfied.

Craciunas~et~al. took the given TTEthernet communication schedule into account while scheduling the tasks on the communication endpoints in~\cite{Craciunas2014}.
They extended the work and introduced the combined network-level and task-level scheduling in~\cite{Craciunas2015}.
The CPU tasks are modeled in a very similar way to the network transmission tasks. 
A CPU is modeled as another physical link in the network.
A task running on this CPU is modeled as a frame which needs to be transmitted over this physical link in one direction, and its transmission time is equal to the required CPU processing time.

We have formulated our problem as the Resource Constrained Project Scheduling Problem (RCPSP~\cite{Neumann2012}), as will be described later, and its Multi-Mode version (MMRCPSP)~\cite{Heller2016}. 
RCPSP is already a well-studied optimization problem.
The survey for the problem and its related variants can be found in~\cite{Hartmann2009}.
The Multi-Mode version, where activities have several alternative modes with different parameters, was studied by Schnell~et~al. in~\cite{SchnellHartl:MMwithPrec}.
The authors developed the exact algorithm by extending the SCIP~Solver~\cite{SCIP}.
Added constraint handlers (i.e. functions for constraints propagation) allowed to directly cooperate between a low-level constraint integer program solver and high-level MMRCPSP constraints and objective.
Another exact approach is described in~\cite{Sitek2016} where the constraint-based modeling tool CP Optimizer was used to solve the RCPSP problem.
The expressive power and universality of the CP Optimizer allows one to easily extend or modify the RCPSP model, which is used for justification of the makespan and porosity optimization in the paper.

\subsection{Paper outline and contribution}
The paper is organized as follows: Section~\ref{Sec:ProblemStatement} describes the studied problem of the TT message scheduling in the TTEthernet network.
In Section~\ref{Sec:Algorithm}, the proposed method is described consisting of a message routing algorithm, a load-balancing heuristic and an RCPSP based formulation of the scheduling problem.
The method is evaluated from the resulting schedule length point of view and also the RC traffic worst-case delay point of view in Section~\ref{Sec:ExperimentalResults}. Section~\ref{Sec:Conclusion} concludes the paper.

The main contribution of the paper is the investigation of a new concept for creating schedules of the TT traffic in the TTEthernet network so that, in opposite to the preceding studies, it focuses on shortening the makespan (the latest completion of any message transmission among all integration cycles and links - see Fig.~\ref{Fig:Link_schedules}) of the TT traffic instead of introducing porosity (blank slots in the TT traffic schedules that are reserved for RC and BE traffic).
The makespan serves as a good measurement for the schedule quality evaluation.
The proposed idea has been inspired by the FlexRay communication scheme, where the Time-Triggered and Event-Triggered segments are separated.
A further contribution is the novel formulation of the TTEthernet scheduling problem as an RCPSP problem. 
Both contributions are evaluated and discussed.

\section{Problem statement}

\label{Sec:ProblemStatement}
This paper aims to design a method for finding feasible periodic schedules for Time-Triggered communication on the TTEthernet network such that the maximal part of the remaining bandwidth can be preserved for RC and BE messages.

\begin{figure}[t]
\resizebox{\columnwidth}{!}
{
\includegraphics{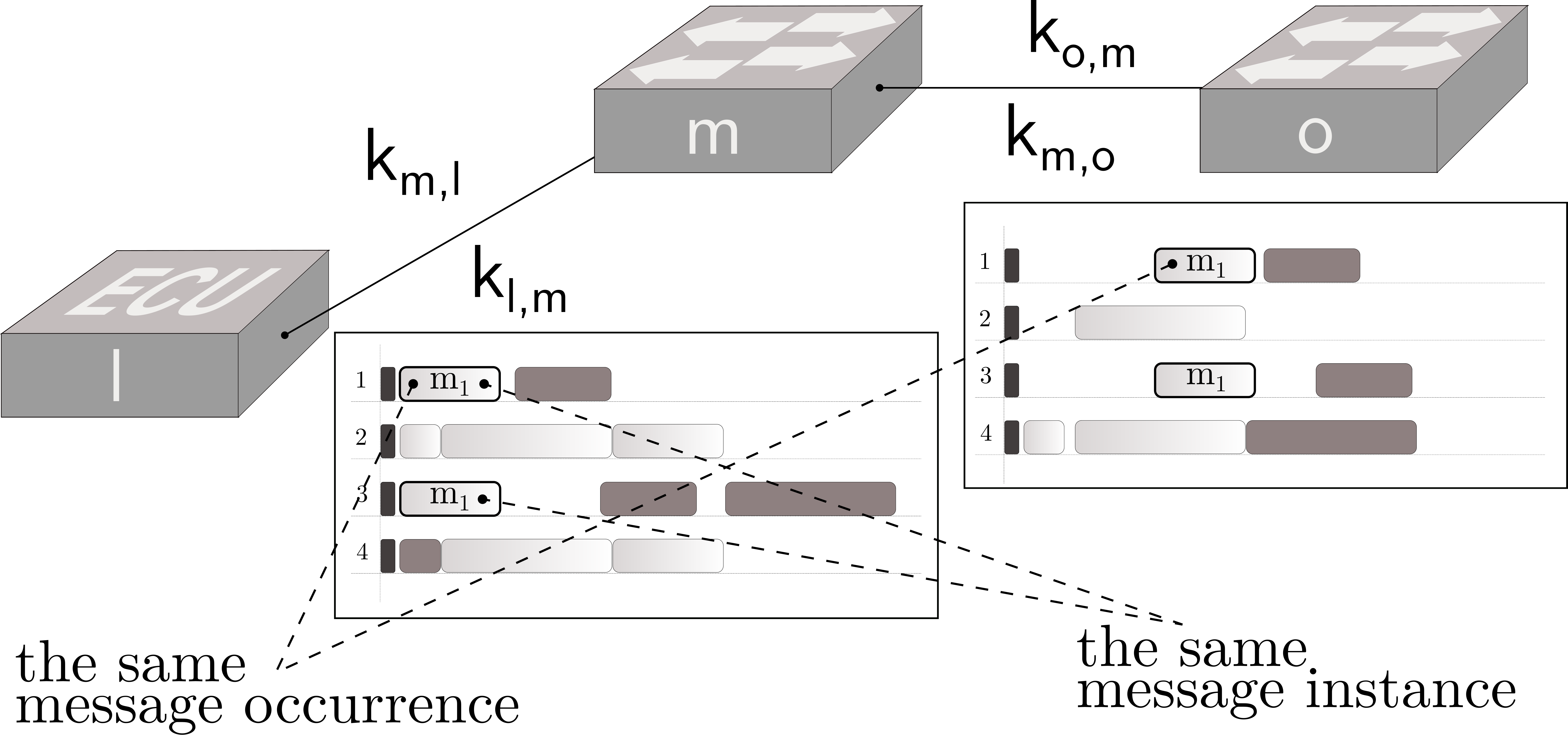}
}
\caption{Visualization of the difference between message occurrence and message instance}
\label{Fig:Messages_terminilogy}
\end{figure}

Each message $m_i$ from a set of the TT messages $M$ has assigned:
\begin{itemize}
\item $t_i$ - period
\item $c_i$ - message length in the number of bits considering headers and interframe gap
\item $d_i$ - deadline
\item $r_i$ - release date
\item $q_i$ - identifier of the transmitting node
\item $Q_i$ - set of the receiving nodes
\end{itemize}
The length of the resulting schedule is determined by the cluster cycle $cc$ (40\,ms in Fig.~\ref{Fig:Link_schedules}). 
The length of the integration cycle $ic$ (10\,ms in Fig.~\ref{Fig:Link_schedules}) is assumed to be the greatest common divisor of message periods (i.e. $ic = gcd_{i | m_i \in M}(t_i)$).
In other words, all the periods $t_i$ has to be an integer multiple of the length of integration cycle $ic$.
The schedule is so-called strictly periodic, which means that the next occurrence of message $m_i$ in a particular link (further called \emph{message occurrence} - see Fig.~\ref{Fig:Messages_terminilogy}) appears in the schedule exactly $t_i$ time units after the current one. 
The positions of all message occurrences of message $m_i$ in the strictly periodic schedule can be deduced from the position of the first message occurrence.

The transmission time of message $m_i$ has to be smaller than or equal to the integration cycle and its length $c_i$ does not exceed the maximal Ethernet frame length of 1530\,bytes.
It would not be possible to send a synchronization message otherwise.
Deadline $d_i$ and release date $r_i$ are assumed to have the value in the range $0 \leq r_i \leq d_i \leq t_i$.

Each node $e_i$ in the network has its identifier assigned.
The nodes are divided into two classes: \emph{redistribution nodes} and \emph{communication endpoints}.
The communication endpoints are nodes that generate or process the data (e.g. sensors, actuators, control units and other ECUs).
Thus, the identifier of any communication endpoint can be assigned to message $m_i$ as transmitter $q_i$ or one of the receivers from set $Q_i$.
The redistribution nodes, on the other side, are switches without any own data to transmit and serve as intermediary nodes for communication.

The TTEthernet infrastructure consists of nodes and links which interconnect them.
Each link $k_{i, j}$ from a set of links $K$ connects two nodes $e_i$ and $e_j$.
This connection covers just one direction of the full-duplex communication.
Therefore, two links $k_{i,j}$ and $k_{j,i}$ model one physical link between nodes $e_i$ and $e_j$.
These two links are independent from the scheduling point of view.

A feasible schedule has to fulfill the following hard constraints:\\
\textbf{Completeness constraint:} Each message $m_i \in M$ has to be scheduled.\\
\textbf{Contention-free constraint:} Any link is capable of transferring at most one message at a time.\\
\textbf{Precedence constraint:}
A sequence of the links $S_i^q = (k_{l,m}, k_{m,o}, ..., k_{p, q})$ represents the path that message $m_i$ has to go from transmitter $q_i = e_l$ to receiver $e_q \in Q_i$ through redistribution nodes $e_m, ..., e_p$. 
An example of such a message transmission is presented in Fig.~\ref{Fig:Infrastructure} where communication endpoints are titled by "ECU" and the redistribution nodes has arrows painted on the top side. The front side of all nodes is labeled by its name.
Only one direction of each physical link is labeled in Fig.~\ref{Fig:Infrastructure} for the sake of simplicity.

The instance of message $m_i$ in link $k_{l,m}$ is denoted as \emph{message instance} $m_i^{l,m}$.
Thus, all the transmissions of some message $m_i$ in one particular link represents the same message instance.
The message occurrence, on the other hand, represents all the transmissions of some message $m_i$ in one particular integration cycle.
The difference between message instance and message occurrence is graphically explained in Fig.~\ref{Fig:Messages_terminilogy}.
The figure shows the detail view on the sub-segment of the network infrastructure from  Fig.~\ref{Fig:Infrastructure} with node $e_i$, $e_m$ and $e_o$ only.
The both links of any physical link are labeled here already.

The path $S_i^q$ is not known in advance if the infrastructure does not have a tree topology.
Therefore, part of the optimization process is to find an appropriate path.
The message transition from transmitting node $q_i$ to all the receivers from $Q_i$ has to be accomplished in one integration cycle.
In the real multi-hop networks, each hop introduces a technical delay caused by queuing in the ingress and egress port, etc.
Such a delay in a switch is represented by parameter $\tau$ for TT messages.
The value of $\tau$ can be in range from \SI{1}{\micro\second} to \SI{2,4}{\micro\second} according to the network configuration~\cite{Steinbach2010}.
Sequence $S_i^q$ entails the generalized precedence constraints because message $m_i$ has to be scheduled in link $k_{m,o}$ $\tau$ time units after it is scheduled in $k_{l,m}$ if $k_{l,m}$ precedes $k_{m,o}$ in $S_i^q$.

The coherent TT traffic segment should be compressed as much as possible to preserve the maximum part of the remaining bandwidth for the RC and BE traffic.
This idea follows the practice from the FlexRay bus or Profinet where the TT traffic has its communication segment.
The TT traffic can be scheduled at the beginning of the integration cycle, and the remaining \emph{TT-free gap} (coherent gap in integration cycle without the TT traffic) is preserved for the RC and BE traffic.
The shortest TT-free gap among all links $k_{i,j}$ is denoted as a \emph{critical gap} (see Fig.~\ref{Fig:Link_schedules}).
Considering the constraints above, the goal of the scheduling is to find a schedule for TT messages which maximizes the critical gap and, thus, minimizes its makespan.

\section{Algorithm}
\label{Sec:Algorithm}
The algorithm proposed in the paper is divided into three stages.
In the first stage (Sec.~\ref{ssec:Routing}), the routing of the messages is established.
In the second stage (Sec.~\ref{SSec:ICAP}), the algorithm finds the assignment of the messages to the particular integration cycles.
The transmission times for each message occurrence in each link are scheduled in the last stage (Sec.~\ref{ssec:scheduling}).

\subsection{Determination of Time-Triggered messages routing}
\label{ssec:Routing}
The network topology is often a tree in industrial networks.
It means that there are no cycles and, therefore, there exists only one possible path from a communication endpoint to any other endpoint.
Thus, the routing determination is trivial in such a case.
However, the TTEthernet does not restrict the network topology, according to the specification, to the tree.
The cycles introduce new redundant paths for messages.
The redundant paths can relieve busy links and serve as a backup during a partial network malfunction.
However, the TT~messages have to know which path they are routed through in advance.
Therefore, the first stage of the algorithm finds the routing.
Each hop the message has to pass implies the increase in overall traffic of the network.
The long path also causes the prolongation of the end-to-end delay because the message is delayed in the ingress and egress buffer of each redistribution node in the path.
This means that utilizing the shortest path algorithm provides an efficient routing with the minimal number of hops.
The network topology is transformed to graph where edges represent the physical links and its weight is set to one.
The Floyd-Warshall algorithm~\cite{Floyd1962} is used, consequently, to find a routing among all nodes in the network.
All the messages follow the routing consequently.

Considering that the aim of this routing is to minimize delay caused by the switch hoping, the routing over the shortest path can cause unbalanced load among links. 
Thus, there are cases where it is better to choose another routing strategy. 
However, this default routing strategy can be substituted by any other routing that follows demands of the particular application.
Other routing strategies can be found for example in \cite{Andrade2016} or \cite{Jouy2014}.

The routing defines the links in which the messages are to be scheduled and specify the precedence relations among message instances of each message.

\subsection{Integration cycle assignment problem}
\label{SSec:ICAP}
The used idea how to distribute messages among the integration cycles comes from the multiprocessor scheduling area.
In the multiprocessor scheduling area, if all the workload of tasks is distributed among the processors evenly, then the schedule makespan has a good chance to be minimal.
Following that, the algorithm tries to distribute the messages among integration cycles evenly.
All the precedence constraints, time lags imposed by the switch delay $\tau$ and real-time constraints are relaxed here.
The integration cycle assignment problem is formulated as the following ILP model:
\begin{equation}
\resizebox{\columnwidth}{!}
{$\displaystyle
\nonumber
\begin{aligned}
& \min_{x_{i,j}}
& & z 
& & & \\
& \text{s.t.}
& & \sum_{j} x_{i,j} = 1
& & & \forall i\\
&
& & \sum_{m_i \in k_{l,m}}{c_i^{l,m} \cdot x_{i,j\bmod t_i}} \leq z
& & & \forall j,l,m\;|\;j \in \{0\,...\,\frac{cc}{ic}\}\\
&
& & x_{i,j} = 0
& & & \forall i,j\;|\;d_i < j\cdot ic \\
&
& & x_{i,j} = 0
& & & \forall i,j\;|\;r_i > j\cdot ic + ic \\
&
& & x_{i,j} \in \{0,1\}; \; z \in R
& & & \forall i,j\\
\end{aligned}
\label{Equ:ILPCLR}
\vspace{0.2cm}
$}
\end{equation}
\addtocounter{equation}{1}
The binary variable $x_{i,j} = 1$ if message $m_i$ is scheduled to the integration cycle $j \in \{0\, ...\, t_i\}$ and zero otherwise.
The parameter $c_i^{l,m}$ represents the transmission time of message $m_i$ on line $l_{l,m}$.
Note, that if the links are configured to have a different bandwidth then the transmission time of the same message varies among the links.
The first constraint assures that the first message occurrence appears in exactly one of the possible integration cycles.
Thus, it satisfies the completeness constraint.
The second constraint makes the variable $z$ to have the value equal to or greater than the time needed to exchange all messages in any integration cycle of any link in the network.
The constraint is evaluated for each resource (each link and each integration cycle in the cluster cycle) such that the transmission time of all message occurrences assigned to the particular resource is summed up.
The resulting total must be less than or equal to variable $z$.
The aim of the ILP model is to find such an assignment that minimizes $z$.
Thus, the maximal time needed for message exchange among all resources is minimized.
The last two constraints force messages to be assigned to the integration cycle which can satisfy the release date and deadline constraints.

The resulting assignment balances the load among the resources, follows the routing of the messages and satisfies the release date and deadline constraints.

\subsection{Link schedule creation problem}
\label{ssec:scheduling}
The RCPSP formulation, inspired by the RCPSP formulation of the traffic scheduling on ProfinetIRT~\cite{Hanzalek2010}, is used for the message scheduling problem.
Thus, let us provide a brief overview for RCPSP first.
\subsubsection{RCPSP Overview}
The RCPSP problem used in the paper is classified as PSm,1,1\textbar temp\textbar $\text{C}_\text{max}$ by Graham's notation.
Translated to common language, the problem is a Project Scheduling problem with \textit{m} resources, one unit of each resource available, and each activity demands at most one unit of the resource.
The problem is constrained by temporal constraints (time lags among activities), and the objective is to minimize the makespan $\text{C}_\text{max}$.

The problem is defined as a sextuplet $(V, p, E, R, B, b)$.
$V = \{A_0, A_1, ..., A_n, A_{n+1}\}$ is a set of non-preemptive \emph{activities} that should be scheduled.
$A_0$ and $A_{n+1}$ are special dummy activities where $A_0$ represents the start of the schedule and $A_{n+1}$ the end of the schedule, respectively.
$p = \{0, p_1, ..., p_n, 0\}$ is the vector of the activities duration. 
The dummy activities have zero duration. 
The resulting schedule would be shifted otherwise.
$E$ is a set of pairs representing time constraints.
If activity $A_i$ has a temporal relation to activity $A_j$ then $(A_i, A_j) \in E$.
Each pair in $E$ is valued by the start-start time lag $l_{i,j}$.
Note that the time lag between any activity $A_i$ and dummy activity $A_{n+1}$ is equal to the activity duration $p_i$.
The temporal constraints are often visualized by the so-called activity-on-node graph $G(V,E)$.
The set $R = \{R_1, ..., R_q\}$ is a set of $q$ resources considered in the problem.
$B$, consequently, represents the number of units available in the resources.
All the resources are unary ($B_k = 1 \;\;\forall k \in R$) in our case.
Parameter $b$ is a set of activity demands where $b_{i,k}$ represents the amount of resource $R_k$ demanded by the execution of activity $A_i$.
The value of $b_{i,k}$ is binary ($b_{i,k} \in \{0, 1\} \;\; \forall i \in V, k \in R$) in the PSm,1,1\textbar temp\textbar $\text{C}_\text{max}$ problem.

The goal is to assign the start time to the activities such that resource demands and time constraints are satisfied and the start time of activity $A_{n+1}$ is minimal.
An interested reader is referred to \cite{Brucker1999} for more detailed information on the RCPSP problem.

\subsubsection{Formulation of the link schedule creation problem to RCPSP}
The stated TTEthernet scheduling problem can be formulated as the PSm,1,1\textbar temp\textbar Cmax problem, considering the following conditions are met:
\begin{enumerate}
\item Routing: It is known in which links each message appears
\item Integration cycle assignment: It is known in which integration cycles each message appears
\end{enumerate}
The first condition is satisfied by the first stage of the algorithm where the messages are assigned to links according to the routing. 
The second condition is also met and the assignment of each message to the integration cycles is obtained from the second stage of the algorithm.

In the RCPSP model~\cite{Neumann2012}, the message instances are represented by the activities.
Thus, the set of message instances is translated to the set of activities $V$ where activity $A_i^{l,m}$ corresponds to message instance $m_i^{l,m}$.
The dummy activities $A_0$ and $A_{n+1}$ are artificially added to $V$.
The duration $p_i^{l,m} = c_i^{l,m}$ of activity $A_i^{l,m}$ expresses the transmission time of message $m_i$ on link $k_{l,m}$.
However, we assume that the bandwidth of all physical links is the same for the sake of simplicity in this paper.
The resulting start time of activity $A_i^{l,m}$ will, consequently, represent the offset of message instance $m_{i}^{l,m}$ in the particular integration cycles in link $k_{l,m}$.
Recall that the integration cycles, in which the message appears, are determined by the first message occurrence.

The set of resources $R$ is obtained from $I$ - the set of integration cycles in the cluster cycle and $K$ - the set of links in the network.
The number of resources used in the model is then $|R| = |I| \cdot |K|$.
Each resource $R_i^{l,m}$ represents the usage of link $k_{l,m}$ in the integration cycle $ic_i$.
The resources have unary availability $B_i^{l,m} = 1$.
Correspondingly, the message demands are unary too, which ensures that the contention-free constraint is satisfied.
Considering that the given conditions are met, the activities demand can be directly derived from the routing and the message to the integration cycles assignment.
The notation $b_{i,j}^{l,m}$, which represents the amount of resource $R_j^{l,m}$ demanded by activity $A_i^{l,m}$, is used from this point further.
If message $m_i$ is to be routed through link $k_{l,m}$, among others, and it is known that the message appears, for example, in $ic_x$, $ic_y$ and $ic_z$ then \mbox{$b_{i,x}^{l,m}=1$}, \mbox{$b_{i,y}^{l,m}=1$}, \mbox{$b_{i,z}^{l,m}=1$}.
All other demands for the activity representing message instance $m_i^{l,m}$ will be equal to zero.

\begin{figure}[t]
\centering
\resizebox{0.65\columnwidth}{!}
{
\includegraphics{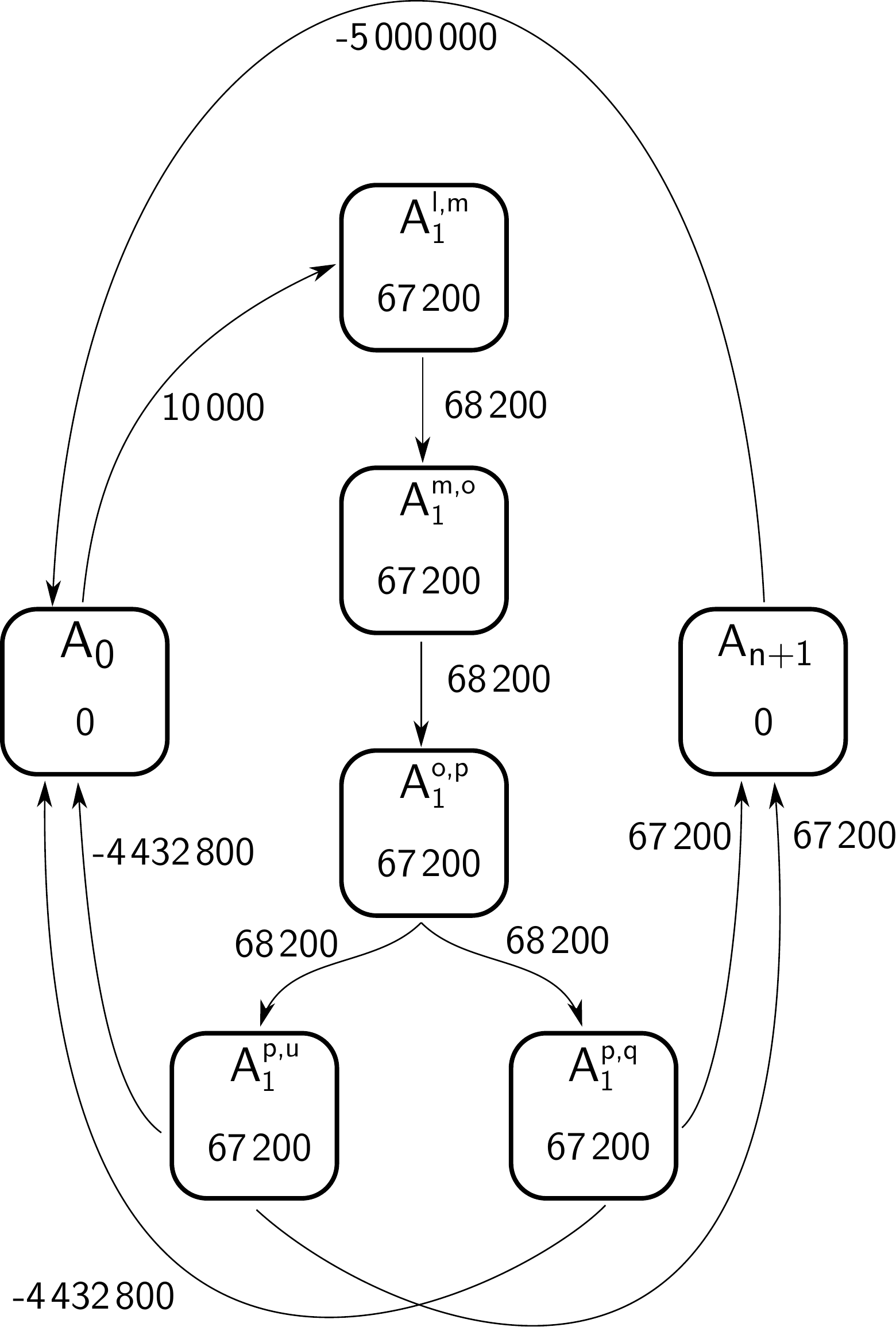}
}
\caption{Example of activity-on-node graph for outlined message $m_1$ from~Fig.~\ref{Fig:Infrastructure}}
\label{Fig:Graph}
\end{figure}

All the real-time constraints are modeled by set $E$ and time lags.
In order to model release date $r_i$ and deadline $d_i$, it is necessary to know their relative values $\tilde{r}_i$ and $\tilde{d}_i$ with respect to the integration cycle in which the message is transmitted.
The values are obtained by subtracting the integration cycle start time from $r_i$ or $d_i$ respectively.
The release date of the message $m_i$ is transformed to the positive time lag led from the dummy activity $A_0$ to activity $A_i^{l,m}$ where $k_{l,m}$ is the first link in the path the message traverses.
The time lag is equal to the relative release date $\tilde{r}_i$.
The deadline, on the other hand, is transformed to a negative time lag led from activity $A_i^{p,q}$, where $k_{p,q}$ is the last link before receiving the communication endpoint, to the dummy activity $A_0$.
The value of the deadline time lag is equal to the negative value of the relative deadline $\tilde{d}_i$ extended by duration $p_i^{p,q}$ (i.e. $-(\tilde{d}_i + p_i^{p,q})$).
Note that if there is more receivers in set $Q_i$ then it is necessary to add the deadline time lag for each receiver separately.
Technical delays and precedence constraints are represented by the positive precedence time lags between two consecutive message instances $m_i^{l,m}$ and $m_i^{m,n}$.
The precedence time lag consists of duration $p_i^{l,m}$ and redistribution nodes delay parameter $\tau$ in store-and-forward networks.

In Fig.~\ref{Fig:Graph}, the example of the activity-on-node graph for message $m_1$ is presented.
Message $m_1$ can be also observed in Fig.~\ref{Fig:Infrastructure} or in its crop in Fig.~\ref{Fig:Messages_terminilogy} - the one with a thick outline.
The links assumed here have a bandwidth of 10 Mbit/s and the message consists of 672 bits (64 octets of Ethernet frame + 12 octets interframe gap + 8 octets preamble and start of frame delimiter).
The message is transmitted by node $e_l$ and received by nodes $e_q$ and $e_u$.
It traverses the redistribution nodes $e_m$, $e_o$ and $e_p$ and the redistribution node delay $\tau$ is \SI{1}{\micro\second}.
The relative release date $\tilde{r}_1$, expressed by the edge from $A_0$ to $A_1^{l,m}$, is \SI{10}{\micro\second}. 
The relative deadline $\tilde{d}_1$, represented by the edge from $A_1^{p,u}$ and $A_1^{p,q}$ to $A_0$, is \SI{4.5}{\milli\second}.

The time lags allow one to limit the latency between the message transmission and reception or the precedence constraints between the messages (e.g. derived from the application level constraints).
However, these constraints are out of the scope of our problem statement and are not used in the paper.

With a given configuration of the RCPSP model, the objective of the RCPSP (i.e. to minimize the maximal makespan) corresponds to the objective of our problem statement.
In other words, the minimal RCPSP makespan assures the maximal critical gap.

The scheduling problem can be formulated as MMRCPSP as well.
In this case, the integration cycle assignment is not necessary to be solved in advance, because it can be formulated as part of the MMRCPSP~\cite{Heller2016}. 
However, the complexity of the resulting MMRCPSP model is significantly higher and, thus, the solver needs more time to find a good or even any solution (see Sec.~\ref{ssec:Quality}).
\section{Experimental results}
\label{Sec:ExperimentalResults}
The proposed scheduling method was tested on a PC with Intel\textsuperscript{ \textregistered} Core\texttrademark\;i7-4610M CPU (two cores with 3~GHz and hyper-threading) and 8~GB RAM. 
The algorithm uses the CPLEX ILP Solver for solving the Integration cycle assignment problem.
The RCPSP problem was solved by both the CP Optimizer and SCIP Solver.
However, the results obtained by the CP Optimizer were significantly better in all the cases.
Thus, the results of the SCIP Solver are omitted here.
\subsection{Benchmark instance sets}
Eight different benchmark sets were used for testing purposes.
Seven of them were generated by our instance generator.
They represent the artificial instances.
The last instance is the real problem instance obtained from our industrial partner.

The synthetic instances are divided into seven groups according to the number of the used TT messages.
The sets are called according to that - from \emph{Set\_20TT} (set with twenty TT messages) to \emph{Set\_2000TT}.
The length of the integration cycle equals thousand times the number of the TT messages, which approximately caused the utilization of half of the bandwidth.
Each such set contains thirty independently generated benchmark instances.
During the instance creation, the algorithm for generating the network topology is selected randomly.
The generator provides four possible algorithms - the star topology generator, the snowflake topology generator, the Barab\'{a}si-Albert algorithm for random tree generation and a random topology generator (the Barab\'{a}si-Albert algorithm with additional redundant links).
The Barab\'{a}si-Albert algorithm is slightly modified.
The nodes of the network with degree 2 are eliminated to remove long linear network segments which are uninteresting from the scheduling point of view.
The network consists of twenty communication endpoints for all the topologies.
Each link has the bandwidth of 1\;Gbit/s.

The messages have a randomly chosen transmitter and set of receivers from the set of communication endpoints.
Thus, any communication endpoint can send a message to an arbitrary subset of other communication endpoints.
The payload of the messages is taken randomly from interval of 46 to 256 bytes.
The message periods should be harmonic or close to harmonic to keep the length of the cluster cycle reasonable.
Thus, the periods are from the domain $t_i = 2^n3^mic$ where $n \in \mathbb{N}$ and $m \in \{0,1\}$.
The release times and deadlines are also generated randomly such that $r_i < d_i \leq t_i$.

The last benchmark set \emph{Set\_Indust} consists of one benchmark instance given to us by our industrial partner. 
The instance evaluated the behavior of the proposed method on instances from a natural environment which often does not directly correspond to the artificial ones.
The network consists of 6 redistribution nodes and 59 communication endpoints and contains alternative paths.
The instance has 1018 messages to exchange with periods from domain $t_i \in \{12.5\text{ms}, 25\text{ms}, 50\text{ms}, 100\text{ms}, 200\text{ms}, 1000\text{ms}\}$ and efficient payloads (payload that contain useful information) from domain $c_i \in [12, 9224]$ bits.
There are no hard real-time constraints imposed on the instance.
Each message has from one up to forty-two receivers.

\subsection{Quality and Performance evaluation of the solver}
\label{ssec:Quality}
The provided method for the TT message scheduling has been evaluated on the described benchmark set and compared with the Multi-mode method.
The Multi-mode method developed by us is similar to the described algorithm, but it uses the Multi-mode RCPSP instead of the general RCPSP model.
In the Multi-mode RCPSP, the alternative modes can be assigned to the activities, and the resulting schedule uses just one of the provided activities.
The TTEthernet problem can be formulated as a Multi-mode RCPSP without the necessity to decide in which integration cycles each message appears in advance.
Thus, it is not necessary to solve the integration cycle assignment problem heuristically by ILP.
The Multi-mode RCPSP can return better solution than our described algorithm denoted as ICAP in Table~\ref{Tab:QualityResults} that uses heuristic integration cycle assignment, but the computational complexity increases significantly.

\begin{table}[t]
\caption{Quality comparison of different approaches}
\centering
\resizebox{\columnwidth}{!}{%
\begin{tabular}{l|r|r|r|r}
\multicolumn{3}{r}{}&\multicolumn{1}{r}{ICAP}&Multi-mode\\
&N [-]&LB [ns]&$\text{C}_{\text{max}}$\;[ns]&$\text{C}_{\text{max}}$\;[ns]\\
\hline
Set\_20TT&124&9626&13430&\textbf{13255}\\
Set\_50TT&312&19468&\textbf{22908}&23678\\
Set\_100TT&624&35603&\textbf{40450}&42291\\
Set\_200TT&1243&66594&\textbf{74663}&79733\\
Set\_500TT&3107&157317&182185&\textbf{181913}\\
Set\_1000TT&6244&318017&\textbf{354577}&354950\\
Set\_2000TT&12460&616446&\textbf{663532}&-\\
Set\_Industrial&5844&59888&\textbf{63568}&-\\
\hline \hline
Average&&177429&194708&-\\
\end{tabular}}
\label{Tab:QualityResults}
\end{table}

Table~\ref{Tab:QualityResults} represents the results obtained by the methods.
The captions of the benchmark sets are situated in the first column.
The second column presents the average number of message instances/activities in the set~(N).
Recall that each message has one message instance for each link the message passes.
Therefore, the number of message instances is often much higher than the number of messages.
All the remaining columns shows the results for the particular algorithm. The values are averaged over thirty independent instances in each set.
The third column shows the lower bound~(LB) for the problem.
The lower bound is obtained from the ILP solution of the integration cycle assignment problem.
The makespan of our problem can never be shorter than the sum of the messages transmission time exchanged in any link if the message to the integration cycle assignment is balanced.
That is why the ILP solution can be used as a lower bound for the whole problem too.
However, it is often not possible to obtain a makespan equal to the lower bound because time lags are not considered in the ILP model.
Therefore, it is not-tight lower bound.
The fourth column contains the average makespan value of the instances in the given set for our provided algorithm~(ICAP) in nanoseconds.
In comparison, the fifth column contains the makespan for the Multi-mode method.
The time limit for the computation of each instance was set to 300 seconds for both methods.

The small theoretical instances with 20 TT messages were solved to an optimum by RCPSP solver by both methods in time.
The final results for these small instances are better for the Multi-mode version because it is not heuristically guided by the ILP solution and, therefore, no states of the search space are pruned.
One can observe the gap between the optimal solution of the Multi-mode method and the non-tight lower bound.
The Multi-mode RCPSP solver was not able to solve bigger instances to an optimum in time.
This leads one to the fact that the guided ICAP method obtains better solutions (except the set with 500 TT messages where relaxing the time-lags in the ICAP formulation led to the weaker integration cycle assignment) from this point further.
Furthermore, no feasible solution was found by the Multi-mode RCPSP solver for instances with two thousand messages in 300 seconds.
Note that the solutions obtained by the ICAP method are not more than 15\% distant from the lower bound for instances with more than 50 TT messages. 

\subsection{Worst-case RC traffic delay evaluation}
To justify the used scheduling method, the solver based on the porosity idea from~\cite{Steiner2011} was implemented to compare the makespan optimization method with the porosity optimization method with respect to the RC traffic.
The porosity is represented by the set of gaps introduced into the schedule.
Consequently, the optimization objective is to maximize the length of these gaps.
Compared to the makespan optimization, which aims to introduce just one big gap at the end of the schedule, the porosity optimization distributes the free bandwidth through the schedule.

Note that unlike in other protocols that has strict separations of time-triggered and event-triggered segments, the introduced porosity gaps as well as one critical gap introduced by makespan optimization does not separate the TT traffic and RC traffic so strictly.
It means that even if the makespan is e.g. 6\,ms (just like in the Fig.~\ref{Fig:Link_schedules}) the RC message can be transmitted sooner if there is no TT traffic scheduled in the particular integration cycle (e.g. message $RC_2$ in Fig.~\ref{Fig:Link_schedules}).
Moreover, the RC traffic can be transmitted even in the gaps of the schedule where they can fit (e.g. message $RC_1$ in Fig.~\ref{Fig:Link_schedules}).

\begin{figure}[h]
\resizebox{\columnwidth}{!}
{
\includegraphics{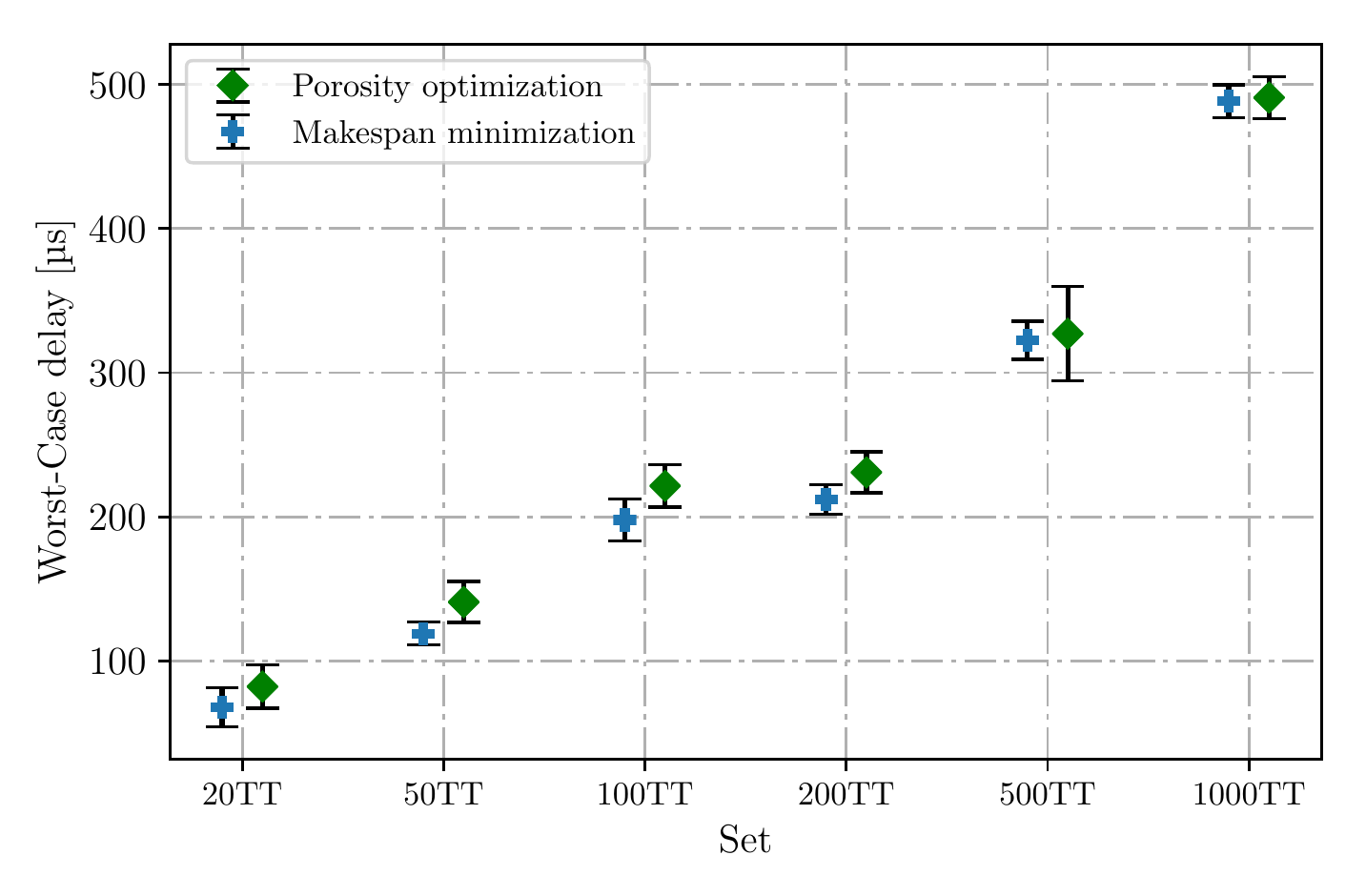}
}
\caption{Worst-Case delays for the RC traffic}
\label{Fig:RC_Results}
\end{figure}

To evaluate the difference between the makespan optimization and the porosity optimization in an objective way, the Worst-Case delay calculation for the RC messages taken from~\cite{TS2015} was employed. 

In Fig.~\ref{Fig:RC_Results}, the average delays among the instances in the particular sets and its standard deviations are presented.
The evaluation was made for all the artificial sets.
However, Set\_2000TT has been omitted from the graph because it caused graph scaling which worsens its clarity.
The worst-case RC traffic delay of Set\_2000TT was $797 \pm 18$\SI{}{\micro\second} for makespan optimization and it was $799 \pm 18$\SI{}{\micro\second} for porosity optimization. 

According to the results, the makespan optimization behaves comparable and even slightly better than the porosity optimization with respect to the RC traffic.
The standard deviation is higher in the case of Set\_500TT because there were a few significant outliers.
Overall, the obtained Worst-Case delays are more stable in the case of the makespan optimization.

\section{Conclusion}
\label{Sec:Conclusion}
The paper in hand focuses on the problem of the Time-Triggered communication scheduling on the TTEthernet network.
To incorporate the Event-Triggered Rate-Constrained traffic with TT traffic is very important for mixed-criticality applications \cite{Novak2016}.
The majority of communication is Event-Triggered in industrial applications nowadays. 
However, the necessity to incorporate safety-related communication, e.g. in automotive or avionics industry, pushes system developers to use the TT traffic because it can be more easily verified and certified.

We have investigated the idea of separating the Time-Triggered traffic and Event-Triggered traffic, which was inspired by the scheme of the FlexRay bus communication cycle.	
The aim has been to minimize the length of the Time-Triggered schedule and leave the rest of the communication bandwidth free for the Rate-Constrained traffic.
For such a problem statement, we have designed the algorithm based on the ILP formulation of the integration cycle assignment problem (load balancing among integration cycles) and the RCPSP formulation of the message scheduling problem. 

The experiments performed on several artificial benchmark sets and one real-case instance show that the designed method is able to obtain near-optimal results (for bigger instances in 15\% of the non-tight lower bound) concerning the proposed criterion.
Moreover, the RC traffic worst-case delay calculations suggest that the proposed concept (minimization of the schedule makespan) can provide slightly better delays in average than the porosity imposed to the schedule by gaps.

For the future work, we would like to study the behavior of the provided method in the environment of the 802.1Qbv~\cite{QbvEthernet} network.
The Ethernet technology is used in the automotive, avionics, etc. industry nowadays, where the development cycle of the products is specific - the new product models are created taking a backward compatibility with the previous ones into account.
Thus, we are also going to study the incremental scheduling case where backward compatibility needs to be preserved.

\section*{Acknowledgment}
This work was supported by the Grant Agency of the Czech Republic under the Project FOREST GACR P103-16-23509S.

\bibliographystyle{ieeetr}
\bibliography{ms}

\end{document}